\documentclass[aps,preprint,showpacs]{revtex4}

\usepackage{graphicx}
\usepackage{hyperref}
\usepackage{amsfonts}
\usepackage{amsbsy}
\usepackage{amssymb}
\usepackage{xcolor}
\usepackage[normalem]{ulem}
\begin{document}

\title{
Statistical characterization of the standard map
}

\author{Guiomar Ruiz$^{1}$}
\email[]{guiomar.ruiz@upm.es}
\author{Ugur Tirnakli$^{2}$}
  \email[]{ugur.tirnakli@ege.edu.tr}
\author{Ernesto P. Borges$^{3,4}$}
  \email[]{ernesto@ufba.br}
  \author{Constantino Tsallis$^{4,5}$}
  \email[]{tsallis@cbpf.br}

\affiliation{
$^1$Departamento de Matem\'{a}tica Aplicada y Estad\'{\i}stica, Universidad Polit\'{e}cnica de Madrid, Pza.\ Cardenal Cisneros s/n, 28040 Madrid, Spain \\
$^2$Department of Physics, Faculty of Science, Ege University, 35100 Izmir, Turkey \\
$^3$Instituto de F\'isica, Universidade Federal da Bahia, Salvador-BA 40170-115 Brazil \\
$^4$National Institute of Science and Technology for Complex Systems \\ \mbox{Rua Xavier Sigaud 150, Rio de Janeiro 22290-180, Brazil}\\
$^5$Centro Brasileiro de Pesquisas Fisicas \\ \mbox{Rua Xavier Sigaud 150, Rio de Janeiro 22290-180, Brazil} }

\date{\today}

\begin{abstract}
The standard map, paradigmatic conservative system in the $(x,p)$ phase space,
has been recently shown [Tirnakli and Borges (2016)] to exhibit interesting
statistical behaviors directly related to the value of the standard map
external parameter $K$.
A detailed numerical description is achieved in the present paper.
More precisely, for large values of $K$ (e.g.,  $K=10$), the Lyapunov exponents
are neatly positive over virtually the entire phase space, and,
consistently with Boltzmann-Gibbs (BG) statistics, we verify
$q_{\text{ent}}=q_{\text{sen}}=q_{\text{stat}}=q_{\text{rel}}=1$,
where $q_{\text{ent}}$ is the $q$-index for which the nonadditive entropy
$S_q \equiv k \frac{1-\sum_{i=1}^W p_i^q}{q-1}$
(with $S_1=S_{BG} \equiv -k\sum_{i=1}^W p_i \ln p_i$)
grows linearly with time before achieving its $W$-dependent saturation value;
$q_{\text{sen}}$ characterizes the time increase of the sensitivity $\xi$
to the initial conditions, i.e.,
$\xi \sim e_{q_{\text{sen}}}^{\lambda_{q_{\text{sen}}} \,t}\;(\lambda_{q_{\text{sen}}}>0)$,
where $e_q^z \equiv[1+(1-q)z]^{1/(1-q)}$; $q_{\text{stat}}$
is the index associated with the $q_{\text{stat}}$-Gaussian distribution
of the time average of successive iterations of  the $x$-coordinate;
finally, $q_{\text{rel}}$ characterizes the $q_{\text{rel}}$-exponential
relaxation with time of the entropy $S_{q_{\text{ent}}}$ towards
its saturation value.
In remarkable contrast, for small values of $K$ (e.g., $K=0.2$),
the Lyapunov exponents are virtually zero over  the entire phase space,
and, consistently with $q$-statistics, we verify
$q_{\text{ent}}=q_{\text{sen}}=0$,
$q_{\text{stat}} \simeq 1.935$,
and $q_{\text{rel}} \simeq1.4$.
The situation corresponding to intermediate values of $K$,
where both stable orbits and a chaotic sea are present, is discussed as well.
The present results transparently illustrate when BG behavior
or $q$-statistical behavior are observed.
\end{abstract}

\pacs{05.10.-a , 05.45.Ac, 05.45.Pq}

\maketitle
\section{Introduction}

Statistical characterization of dynamical systems,
particularly weakly chaotic low-dimensional dissipative maps,
have been made according to
its sensitivity to the initial conditions,
the rate of increase of the entropy,
the rate of relaxation to the attractor,
and the distribution of a statistical variable
\cite{Tsallis2009}.
For strongly chaotic systems, these behaviors present an exponential nature
and are described by the Boltzmann-Gibbs (BG) statistical mechanical framework
\cite{BeckSchlogl}.
Weakly chaotic systems, that present zero {\it largest Lyapunov exponent} (LLE)
$\lambda$,
need a more general formalism to be properly handled.
It has been observed that the nonextensive statistical mechanical framework
is able to describe different features of various classes of complex systems
\cite{Rapisarda-2004, Esquivel-2010, CMS-2010, Shao-2010, Sotolongo-2010, Mohanalin-2010, ALICE-2010, PHENIXbis, PHENIX-2011,ALICE-2011,ATLAS-2011,Wong-2012,Marques-2013,Bogachev-2014, Combe-2015},
including the logistic map \cite{Baldovin-Robledo-2002}.
If $\lambda$ approaches zero,
the sensitivity to the initial conditions
and the rate of relaxation to the attractor
follow $q$-exponential behaviors,
the distributions follow a $q$-Gaussian form,
and the rate of increase of entropy demands $S_q$ entropy,
instead of BG entropy.
All these behaviors are parameterized by the indices
$q_{\text{sen}}$,
$q_{\text{ent}}$,
$q_{\text{rel}}$,
$q_{\text{stat}}$,
for {\it sensitivity},
{\it entropy production},
{\it relaxation},
and
{\it stationary distribution},
respectively.
It has been observed (and in some instances proved) that,
for low-dimensional systems, $q_{\text{sen}}= q_{\text{ent}}$,
through the $q$-generalized Pesin equality \cite{Tsallis2009}.
Consequently, three (possibly related) quantities remain to be addressed,
that are currently referred to as the $q$-triplet \cite{Tsallis2009}.
The $q$-triplet has been experimentally observed in a variety of complex
systems like in the ozone layer \cite{Ferri2010} and in the
Voyager I data for the  solar  wind \cite{Burlaga2013}, among many others.
Relations are expected to connect
$q_{\text{ent}}$, $q_{\text{stat}}$, and $q_{\text{rel}}$,
ultimately completely determined
by the nonlinear dynamics of the system.
The possible relations between the $q$-triplet indices
should in principle exhibit the fixed point $q=1$ (BG statistics).
At least for low-dimensional systems, if one of them equals unity,
the others also equal unity.  This typically happens whenever $\lambda > 0$.
In fact, LLE plays a central role in the definition of which statistical
framework is to be used in the description of nonlinear dynamical systems.

In the present work we address a low-dimensional conservative system,
namely the standard map \cite{chirikov,zaslavsky1}:
\begin{equation}
 \label{eq:stdmap}
 p_{i+1} = p_i - K \sin x_i\qquad;\qquad x_{i+1}=x_i+p_{i+1}
\end{equation}
($p$ and $x$ are taken as modulo $2\pi$).
Its phase space can exhibit strongly chaotic regions,
characterized by positive local LLE $\lambda$,
and weakly chaotic regions, with zero values of $\lambda$.
The LLE has been numerically evaluated through
the Benettin et al algorithm \cite{benettin}.
It is calculated for each initial condition separately,
as illustrated by Fig.\ 2 of \cite{tirnakli-borges-2016}
(see also \cite{BarangerLatoraRapisarda2002}).
The extent of the phase space for each case (strongly or weakly chaotic)
is determined by the control parameter $K$.
For sufficiently small values of $K$, for instance $K=0.2$,
the entire phase space is practically stable (zero LLE),
and if $K$ takes high values ($K=10$), the points of the phase space
have finite positive values of $\lambda$.
In the present work, we evaluate the values of $q_{\text{sen}}$,
$q_{\text{ent}}$, and $q_{\text{rel}}$
for both strongly and weakly chaotic cases.
We also revisit the stationary distributions, particularly the mixed case
(strongly and weakly chaotic),
that has been recently found for this system \cite{tirnakli-borges-2016}.
Finally, we have determined the relation of strongly and weakly chaotic
$q_{\text{stat}}$  indices, with respect to the stationary distributions
of the map for arbitrary values of $K$.

\section{\label{sec:q-sen} Sensitivity to initial conditions}

The sensitivity to initial conditions is given by
\begin{equation}
 \label{eq:sensitiv}
  \xi(t)=\lim_{\|{\bf \Delta x}(0)\|  \to {\bf 0}}
         \frac{\|{\bf \Delta x}(t)\|}{\|{\bf \Delta x}(0)\|}
\end{equation}
where ${\bf\Delta x}(t)$ is the temporal dependence of the discrepancy
of two very close initial conditions at time $t$.
When ergodic behavior dominates over the whole phase space,
Eq.\ (\ref{eq:sensitiv}) is expressed by the exponential dependence,
\begin{equation}
 \label{eq:xi-exp}
 \xi(t) = e^{\lambda t}
\end{equation}
and $\lambda$ is the standard maximum Lyapunov exponent.
Strongly chaotic systems are those which present $\lambda>0$.
They are strongly sensitive to initial conditions.
Insensitivity to initial conditions are described by $\lambda <0$.
The marginal case $\lambda =0$ is governed  by a more subtle rule
than
Eq.\ (\ref{eq:xi-exp})
\cite{tsallis-plastino-zheng-1997}:
\begin{equation}
 \label{eq:xi-q-exp}
 \xi(t) = \exp_{q_{\text{sen}}} (\lambda_{q_{\text{sen}}} t)
\end{equation}
where $\exp_q x \equiv [1 + (1-q) x]_+^{1/(1-q)}$,
$[u]_+=\max\{u,0\}$, is the so called  $q$--exponential function
\cite{tsallis:quimica_nova},
and $\lambda_{q_{\text{sen}}}$ is a generalized Lyapunov exponent.
If $\lambda_{q_{\text{sen}}} > 0$  (with $q_{\text{sen}}<1$)
the system is weakly chaotic, in the sense that
$\xi(t)$
diverges slower than the exponential (strongly chaotic) case,
--- asymptotically, a power law.
If the system is strongly chaotic,
$q_{\text{sen}} \to 1$,
and Eq.\ (\ref{eq:xi-q-exp}) recovers the standard exponential dependence,
with $\lambda_{q_{\text{sen}}}\to \lambda_1 \equiv \lambda$.

\begin{figure}[h]
\centering
 \includegraphics[width=0.4\columnwidth,keepaspectratio,clip=]{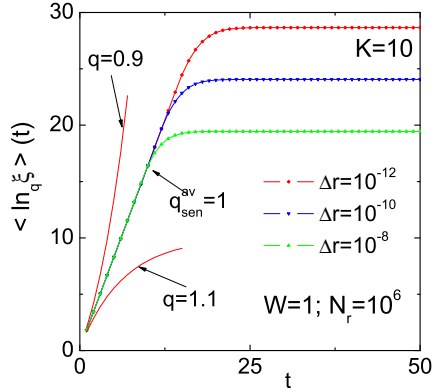}
  \vspace{-0.5cm}\caption{\label{fig:lnqzik10b}
(Color online) $\langle \ln_q \xi (t)\rangle$ vs.\ $t$, for $K=10$.
The $N_r=10^6$ realizations  are evaluated over the whole (not partitioned)
phase space, i.e. $W=1$ (see text).
Decreasing Euclidean distances between initial conditions
$\Delta r\equiv\| {\bf \Delta x}(0)\|$ are considered.
Linear dependence of $\langle \ln_{q_{\text{sen}}} \xi \rangle$ with time
is yielded for $q^{\text{av}}_{\text{sen}}=1$.
}
\end{figure}

The correctness of this description can be checked, for both chaotic regimes
of standard map, in Figures \ref{fig:lnqzik10b} and \ref{fig:lnqzik02}.
They show, for different values of $q$, the averages of $\ln_q\xi (t)$
(where $\ln_qx\equiv (x^{1-q}-1)/(1-q)$ is the inverse function of the
$q$-exponential, and $\ln_1 x=\ln x$)
over $N_r$ realizations.   Each  realization starts with a randomly chosen
pair of very close initial conditions, that are localized inside a
particular  cell of the $W$ equally partitioned phase space.
In Fig.\ \ref{fig:lnqzik10b}, that corresponds to the case of strong chaos
$K=10$, where LLE $>0$), we consider decreasing initial discrepancies in
Eq.\ (\ref{eq:sensitiv}) so as to obtain a well defined behavior for
increasingly long times.
In all cases (strongly and weakly chaotic chaos), after a transient,
we verify a nontrivial property \cite{ananos-tsallis-2004}
namely that exists a special value of $q$, noted $q^{\text{av}}_{\text{sen}}$
(where  av stands  for  {\it average}),
which yields a linear dependence of
$\langle \ln_{q_{\text{sen}}} \xi \rangle$
with time.
In other words,   we verify
$\langle \ln_q \xi(t)\rangle \approx \lambda_{q_{\text{sen}}^{\text{av}}} t$.

\begin{figure}[h]
\centering
 \hspace{-0.5cm}
 \includegraphics[width=0.54\columnwidth,keepaspectratio,clip=]{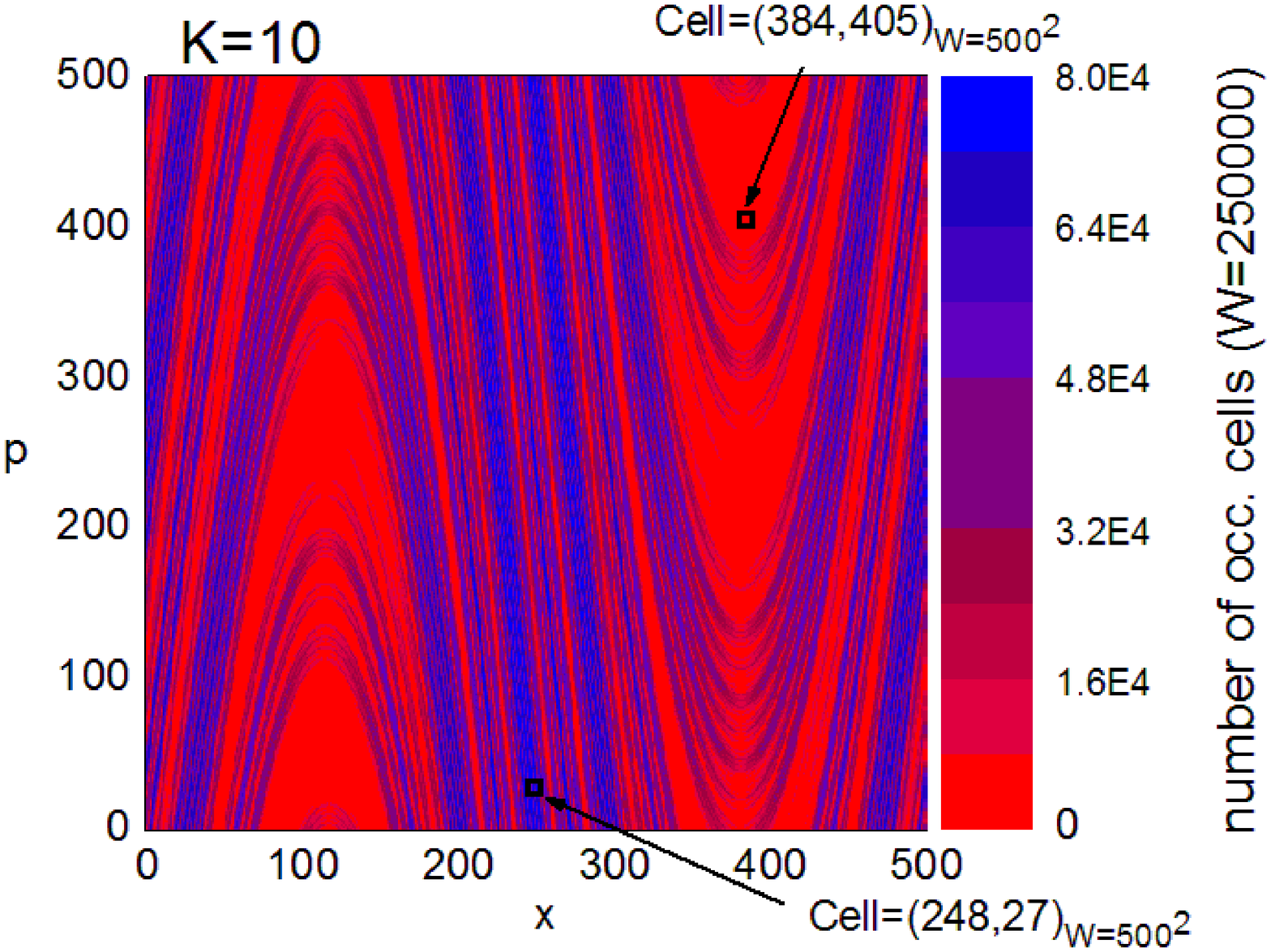}
 \includegraphics[width=0.46\columnwidth,keepaspectratio,clip=]{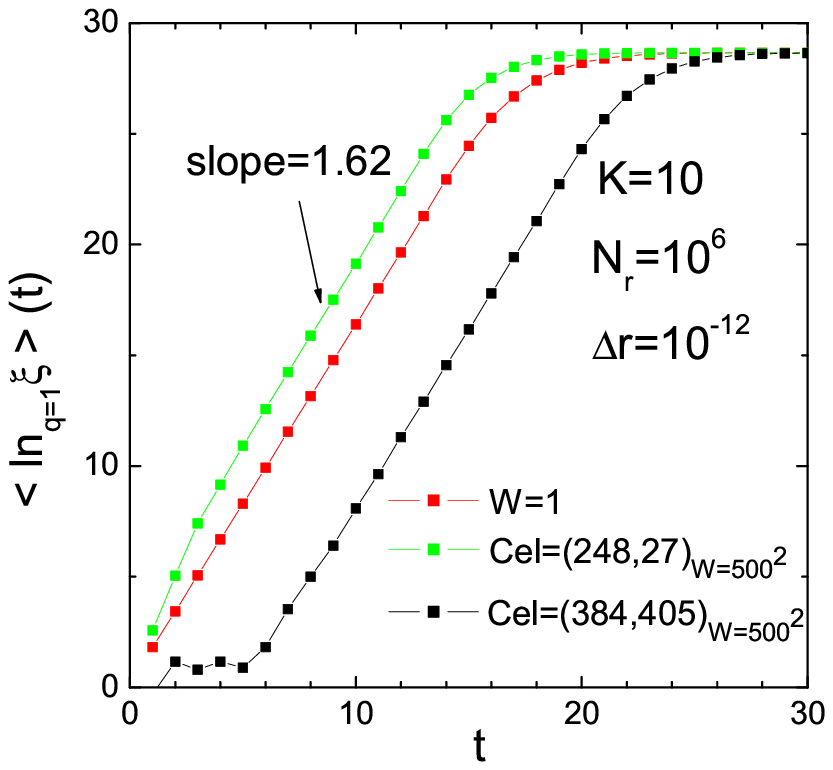}
 \hspace{-0.5cm} \vspace{-1cm}\caption{\label{fig:lnqzik10a}
(Color online)
{\bf Left:} Color map of the integrated number of occupied cells
in a $W=500\times 500 $ equally partitioned phase space,
from iteration $1$ to iteration $10$, as a function of the
$(x^{\text{th}},p^{\text{th}})$ localization of the cell that
contains initial conditions.
{\bf Right:} Temporal evolution of $\langle \ln_{q=1} \xi \rangle$
for $K=10$, where $\Delta r\equiv\| {\bf \Delta x}(0)\|=10^{-12}$
is the Euclidean distance of the $N_r$  pairs of initial conditions.
The $N_r=10^6$ realizations  are estimated over one of
the quickest spreading cells
(the cell at line $248$ and column $27$, green squares)
and over one of  the  slowest  spreading cells
(the cell at line $384$ and column $405$, black squares),
amongst the $W=500\times 500$ cells of an equally partitioned phase
space. Averaged temporal evolution $\langle \ln_{q=1} \xi (t)\rangle$ over the
not partitioned phase space is also shown ($W=1$, red squares).
}
\end{figure}

For the strongly chaotic case, exponential
sensitivity to the initial conditions is verified with
$q_{\text{sen}}^{\text{av}}=1$, and the
same generalized Lyapunov  exponent characterizes the whole phase space no mater
the value of $W$ or the particular cell to be chosen.  Fig.\ \ref{fig:lnqzik10a}
exemplifies this fact, as it shows that the slope of the temporal
evolution of the averaged sensitivity to initial conditions is preserved over an
intermediate regime where correlation coefficient equals one and
$\lambda=\lambda_{q_{\text{sen}}^{\text{av}}=1}=1.62$ (time steps)$^{-1}$,
in accordance with \cite{latora-baranger1999}.
Consequently, generalized Lyapunov exponent does not depend on the spreading
occupation index \cite{latora-baranger-tsallis-rapisarda-2000}  of averaging
cells, and it makes sense choosing the random pairs of initial conditions
to compute the  $N_r$  realizations all over the phase space, i.e. $W=1$.

\begin{figure}[ht]
\centering
\hspace{-0.5cm}
 \includegraphics[width=0.52\columnwidth,keepaspectratio,clip=]{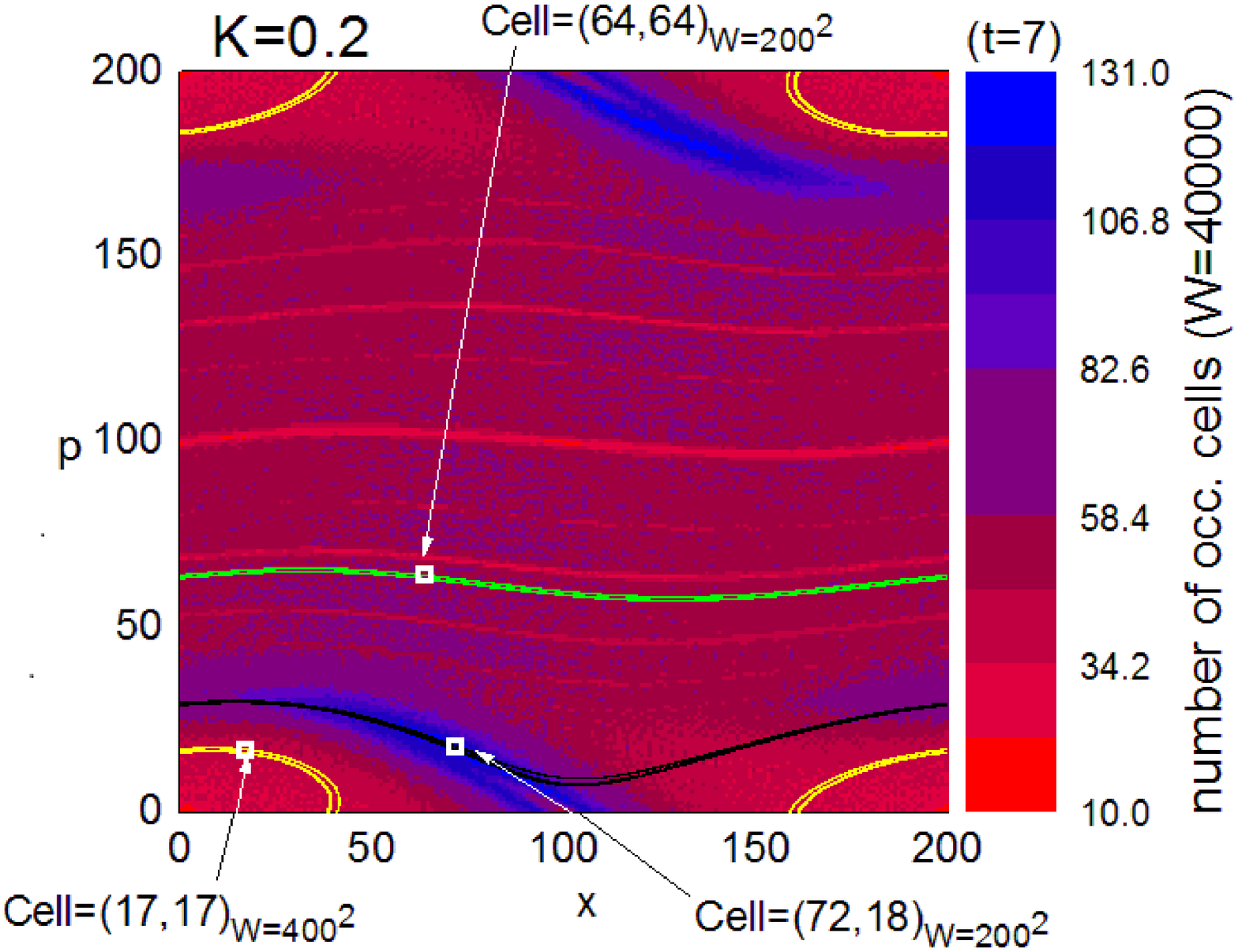}
\hspace{-1.5cm}
  \includegraphics[width=0.50\columnwidth,keepaspectratio,clip=]{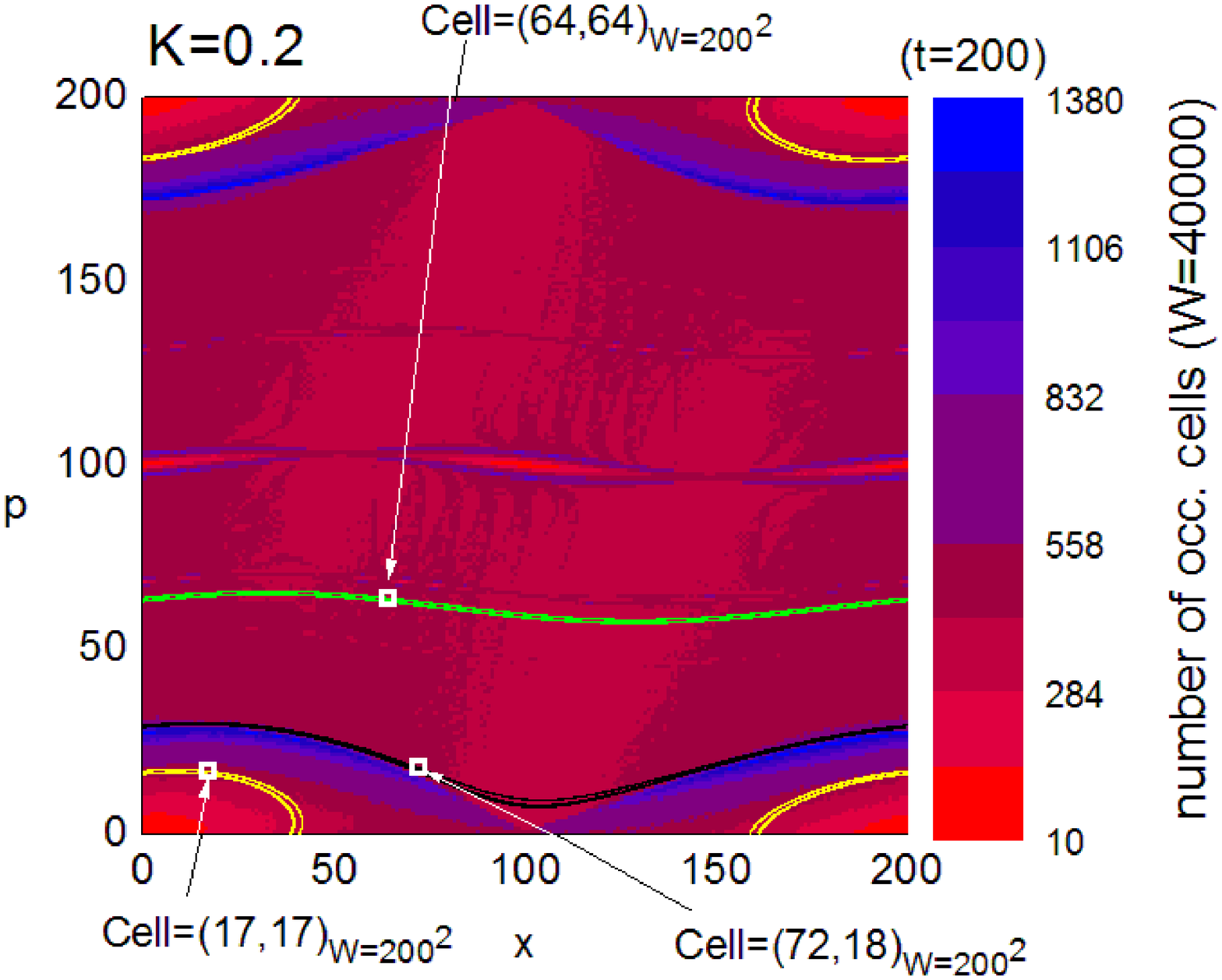}
 \caption{\label{fig:OcColK02W400}
(Color online)
Color maps of the integrated number of occupied cells in a $W=200\times200$
equally partitioned phase space, from iteration $1$ to iteration
$t$ ($t=7$, left,  $t=200$, right), as a function of the
$(x^{\text{th}},p^{\text{th}})$ localization of the cell that contains
the initial conditions.
A set of quasi-periodic trajectories are superimposed to the maps,
in order to warn about the different local transient
times (yellow, black and green lines).
}
\end{figure}

\begin{figure}[ht]
\centering
 \hspace{-0.5cm}
 \includegraphics[width=0.5\columnwidth,keepaspectratio,clip=]{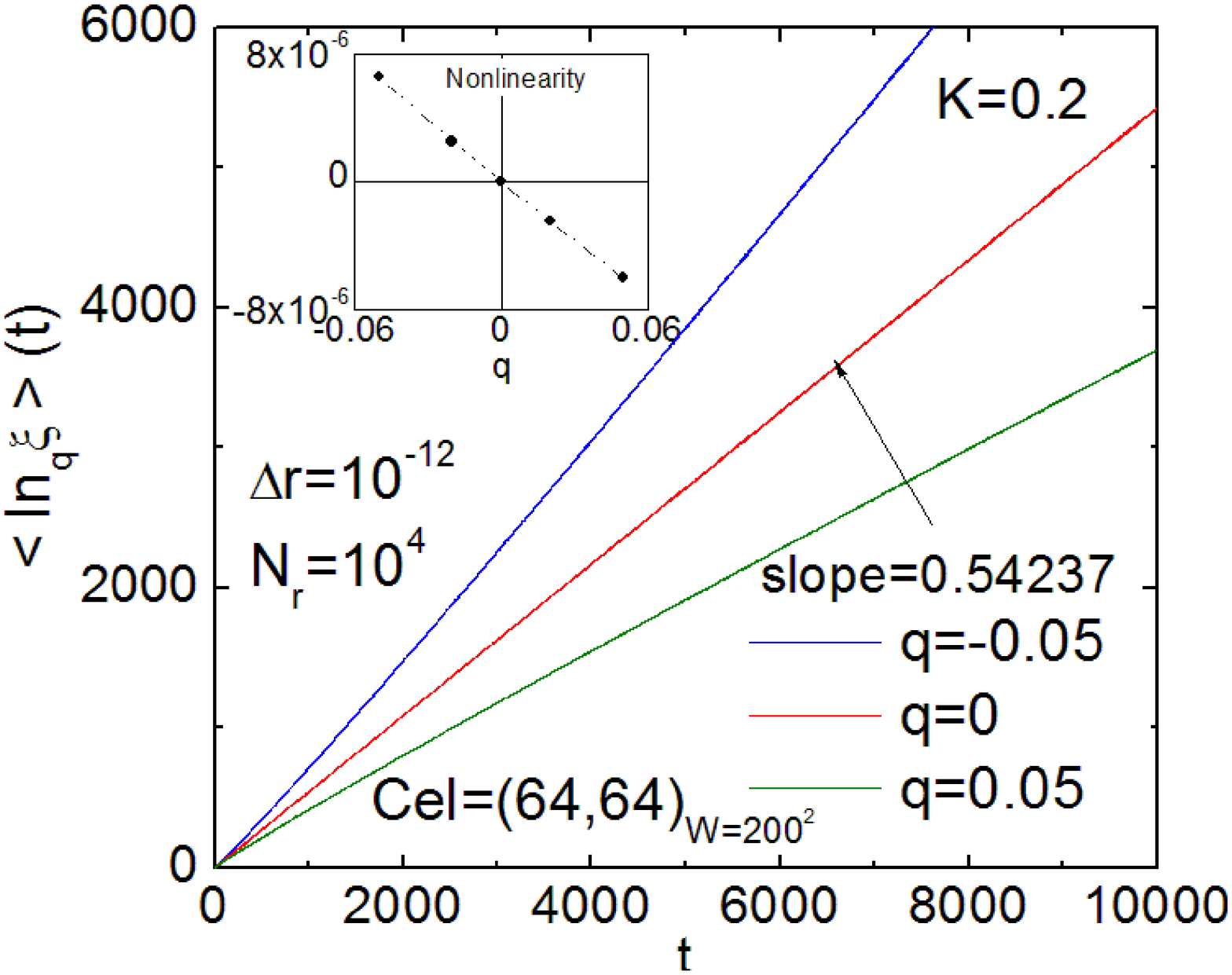}
 \hspace{-0.5cm}
 \includegraphics[width=0.5\columnwidth,keepaspectratio,clip=]{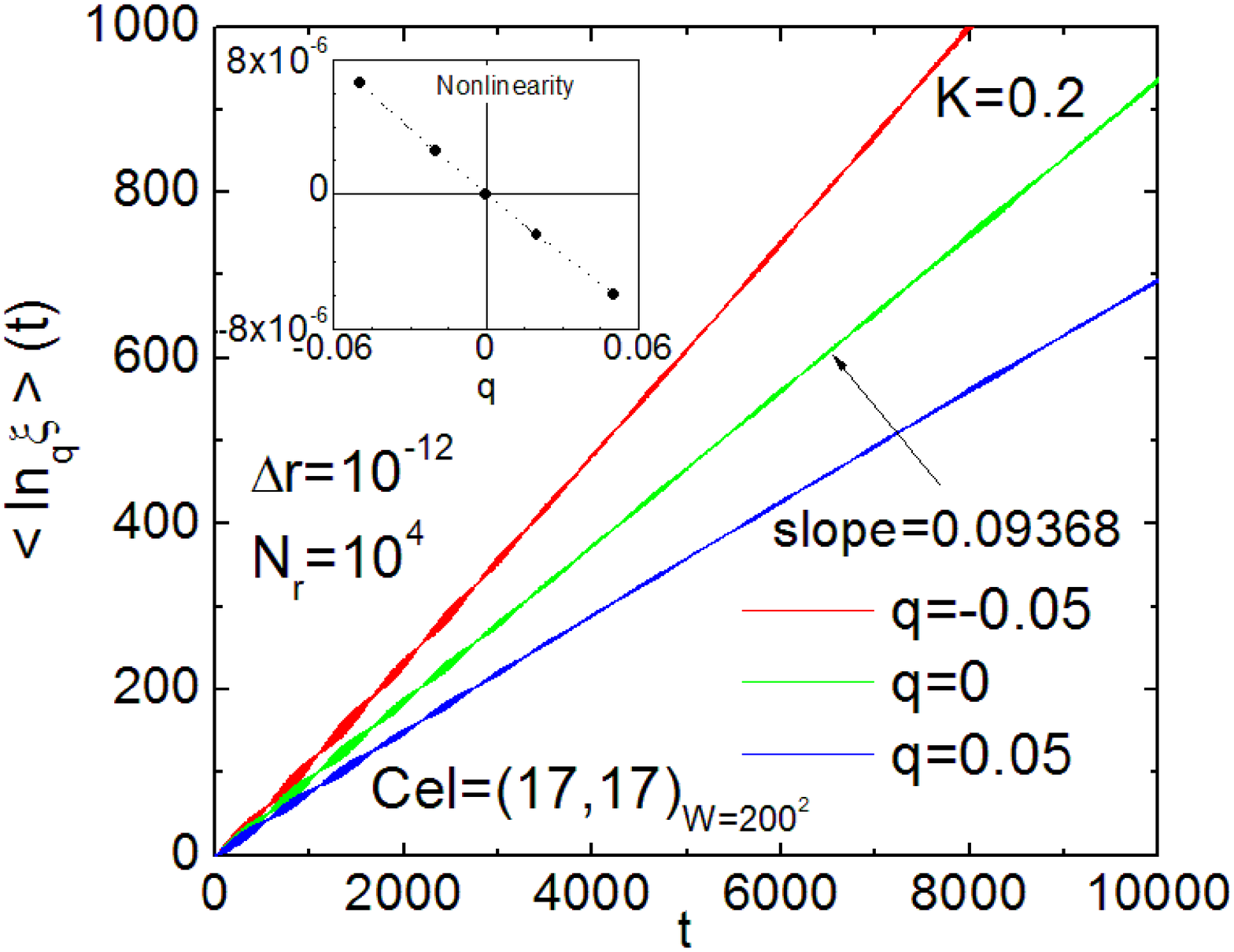}
\caption{\label{fig:lnqzik02}
(Color online) Temporal evolution of $\langle \ln \xi_q \rangle$,
when  $K=0.2$, for $(64^{\text{th}},64^{\text{th}})$ (left figure)
and  $(17^{\text{th}},17^{\text{th}})$ (right figure)
initial cells in the $W=200\times 200$ equally partitioned phase space.
Averages are estimated for a set of $N_r=10^4$ realizations,
whose initial conditions have been randomly chosen in their respective
initial cells.
The $(17^{\text{th}},17^{\text{th}})$  case presents a slightly periodically
modulated evolution that is due to the not connected character of trajectories
(see  Fig.\ \ref{fig:OcColK02W400}).
Insets represent the nonlinearity measure $R\equiv C/B$ in the fitting curve
of $\langle \ln \xi_q (t)\rangle$, $A+Bt+Ct^2$,
and demonstrate that $q_{\text{sen}}^{\text{av}}=0$ is the optimum value
of $q$ (a straight line fitting curve is obtained).}
\end{figure}

\begin{figure}[ht]
\centering
 \hspace{-0.5cm}
 \includegraphics[width=0.5\columnwidth,keepaspectratio,clip=]{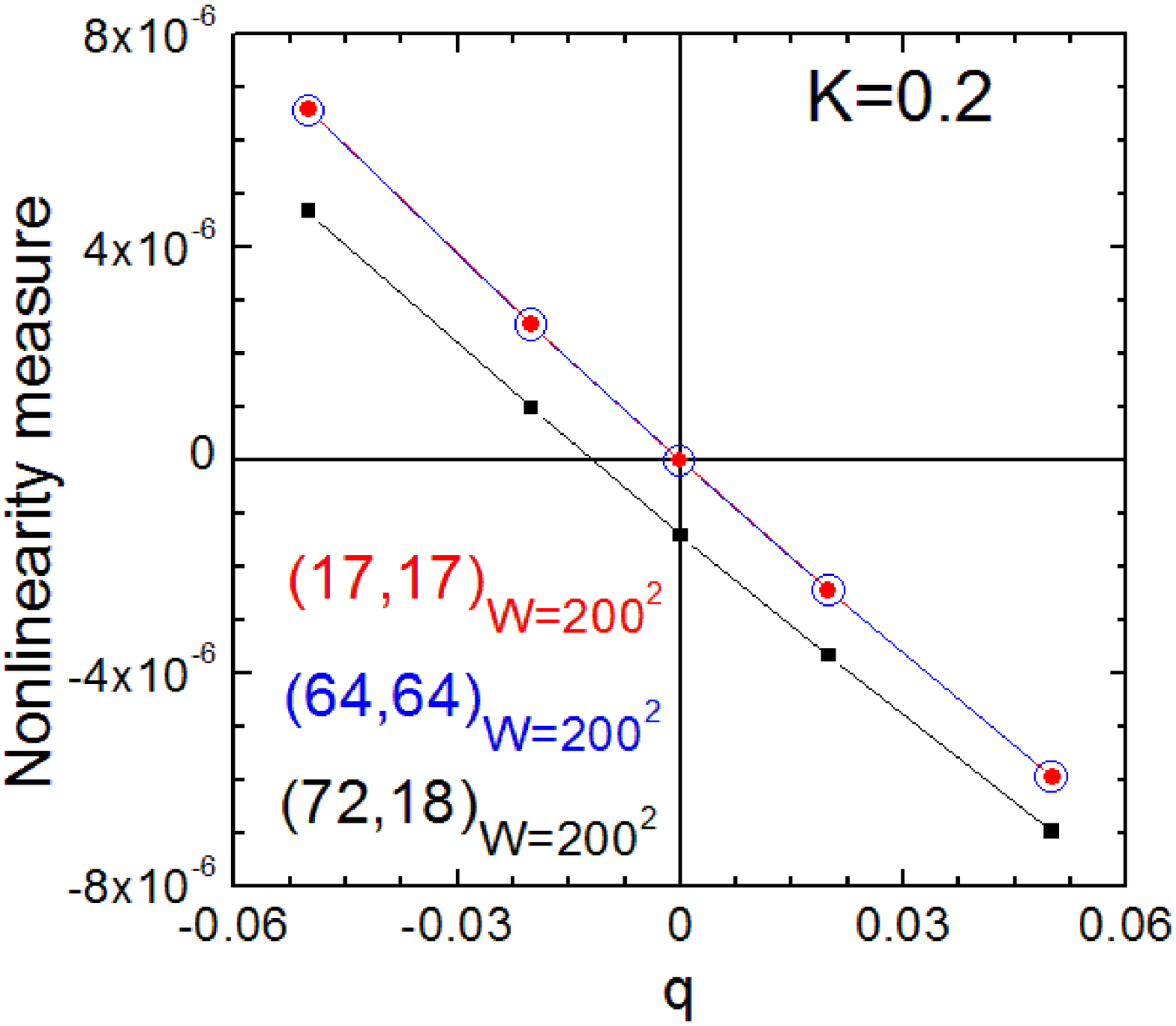}
 \hspace{-0.5cm}
 \caption{\label{fig:lnqzik02qneg}
(Color online) Nonlinearity measure of the polynomial fitting curve
$\langle \ln \xi_q (t)\rangle$  vs.\ $q$, for averages over
$(17^{\text{th}},17^{\text{th}})$,  $(64^{\text{th}},64^{\text{th}})$
and  $(72^{\text{th}},18^{\text{th}})$ cells of the $W=200\times 200$
equally partitioned phase space of the standard map with $K=0.2$.
Red and blue lines are superimposed.
         }
\end{figure}

For the weakly chaotic case,
stability islands dominate the phase space.
Consequently, the characterization of the spreading occupation index of cells
is much more intricate.
In this case, a ``global characterization'' of the map has tentatively been made
\cite{Baldovin-2002, Baldovin-2004}.
But Fig.\ \ref{fig:OcColK02W400} ($K=0.2$, where LLE = $0$) demonstrates that a
common transient time can not be defined, and this fact tends to preclude a neat
linear dependence of $\langle \ln_{q_{\text{sen}}} \xi \rangle$ with time.
We now do not choose the $N_r$ random pairs of initial conditions
all over the phase space, but in particular cells, i.e., in local regions.
Nevertheless, we identify regions that exhibit $q_{\text{sen}}^{\text{av}}=0$,
whose value of $\lambda_{q_{\text{sen}}^{\text{av}}}$ depend on the
initial cell, as shown in Figure \ref{fig:lnqzik02}.
In fact, a huge set of randomly chosen cells exhibit analogous results of
$q_{\text{sen}}^{\text{av}}=0$,
but we have also found cells where linear behavior is satisfied for
$q_{\text{sen}}^{\text{av}}\ne 0$ (Fig.\ \ref{fig:lnqzik02qneg}).

\section{\label{sec:q-ent} The rate of entropy production}

%

With respect to the entropy production per unit time,
we may conveniently use the $q$-entropy ($k=1$, henceforth)
\cite{Tsallis1988,Tsallis2009}

\begin{equation}
S_q(t)= k \frac{1-\sum_{i=1}^W p_i^q}{q-1} \qquad (S_1=S_{BG}=-k\sum_{i=1}^W p_i \ln p_i)
.
\label{eq:qentropy}
\end{equation}

The $q$-entropy production is estimated dividing the phase space in $W$ equal
cells, and randomly choosing $N_{\text{ic}} \gg W$  initial conditions inside one
of the $W$ cells (typically $N_{\text{ic}}=10W$).
We follow the spreading of points within the phase space, and
calculate Eq.\ (\ref{eq:qentropy}) from the set of occupancy probabilities
$\{p_i(t)\}$ ($i=1,2,\dots,W$).
Fluctuations are of course present, but they can be reduced
when we choose $N_c\gg 1$ initial cells
within which the $N_{\text{ic}}$ initial conditions are chosen,
and  average $S_q(t)$ over the $N_{c}$ realizations.
The proper value of the entropic parameter $q_{\text{ent}}^{\text{av}}$
is the special value of $q$ which makes the averaged $q$-entropy
production per unit time to be finite,
which is ultimately related with the extensivity of the entropy.
The $q$-entropy production per unit time is consequently obtained as
\begin{equation}
K_{q_{\text{ent}}^{\text{av}}}=
      \lim_{t \to \infty } \lim_{W \to \infty } \lim_{N_{\text{ic}} \to \infty }
      \frac{\langle S_{q_{\text{ent}}^{\text{av}}}(t)\rangle_{N_{c}}}{t}<\infty,
\label{eq:qKolmogorov}
\end{equation}
and must be calculated taking into account that the partitions of phase space,
$N_c$ and $N_{\text{ic}}$ are such as to obtain robust results.

\begin{figure}[ht]
\centering
 \includegraphics[width=0.6\columnwidth,keepaspectratio,clip=]{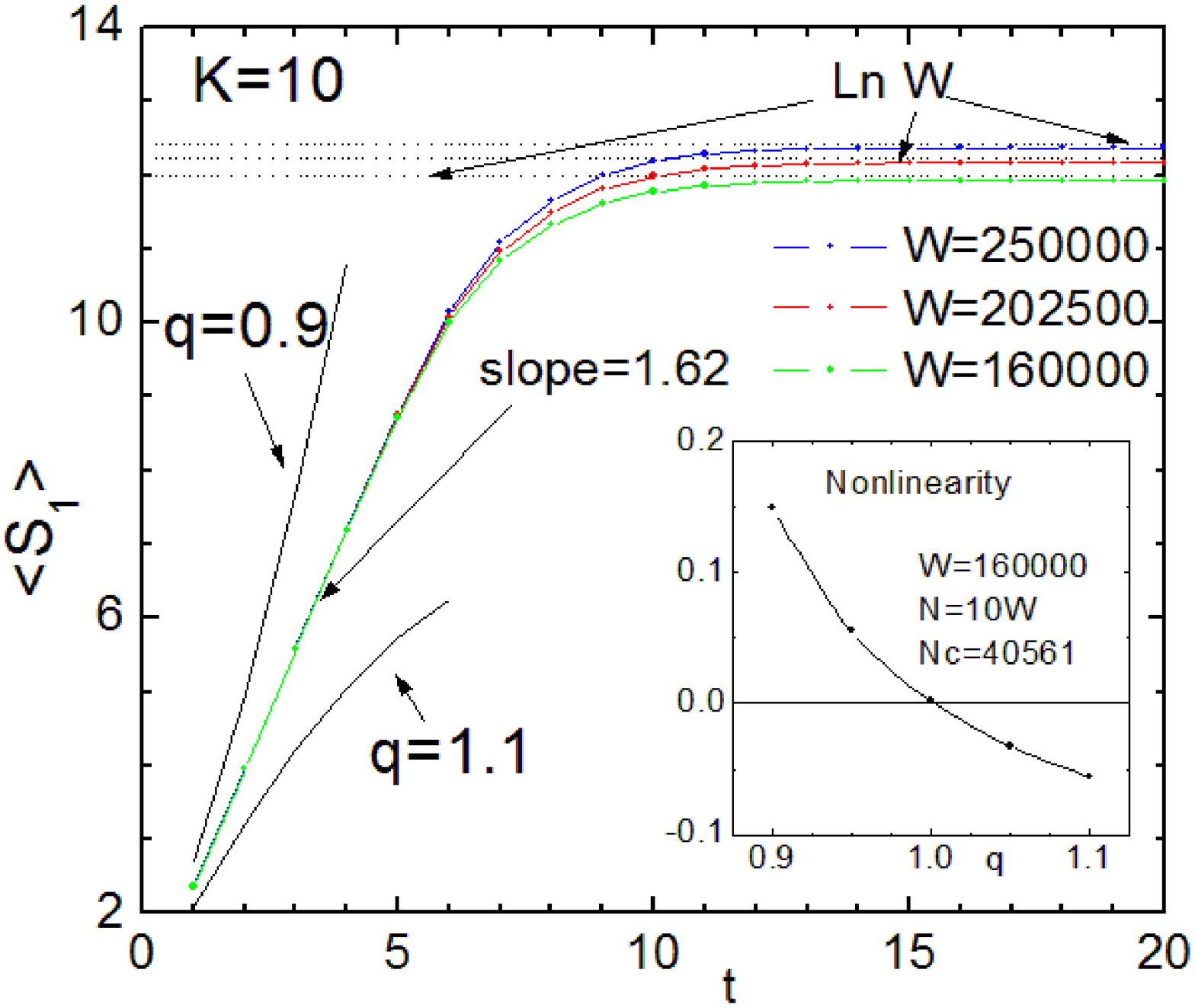}
 \caption{\label{fig:s1k10}
(Color online) Temporal dependence of $(q_{\text{ent}}^{\text{av}}=1)$-entropy
in the standard map with $K=10$,
averaging over $1\% W$, for different values
of the number of cells in the equally partitioned phase space, $W$.
The case of $q<q_{\text{ent}}^{\text{av}}=1$
($q>q_{\text{ent}}^{\text{av}}=1$)
corresponds to positive (negative) concavity.
The $t\to \infty $ limiting value is $\langle S_1 \rangle =\ln W$.  }
\end{figure}

When the LLE is largely positive
and the phase space is dominated by a chaotic sea,
the Boltzmann-Gibbs entropy ($S_1 = S_{BG}\equiv - \sum_{i=1}^W p_i \ln p_i$)
is expected to be the appropriate one (i.e. $q_{\text{ent}}=1$).
Our numerical results show, as expected, that the proper value
of the entropic index is
$q_{\text{ent}}^{\text{av}}=q_{\text{sen}}^{\text{av}}=1$ for $K=10$
(see Figure \ref{fig:s1k10}). Both indexes have been obtained comparing
the nonlinearity measure of the polynomial fitting curves
over the intermediate regime (after a transient and before saturation)
were the correlation coefficient is constant and equals $1$.
Another interesting result that emerged is the coincidence of the slopes of
the sensitivity and entropy functions of time
(see Fig.~\ref{fig:lnqzik10a} and Fig.~\ref{fig:s1k10}),
i.e., the $q$-entropy production per unit time satisfies
$K_{q_{\text{sen}}^{\text{av}}=1}=\lambda_{q_{\text{sen}}^{\text{av}}=1}$.
These results are numerically compatible with a $q$-generalization of a
Pesin-like identity for ensemble averages in strongly chaotic conservative maps,
and reinforce those in \cite{ananos-tsallis-2004}.

\begin{figure}[ht]
\center
 \includegraphics[width=0.5\columnwidth,keepaspectratio,clip=]{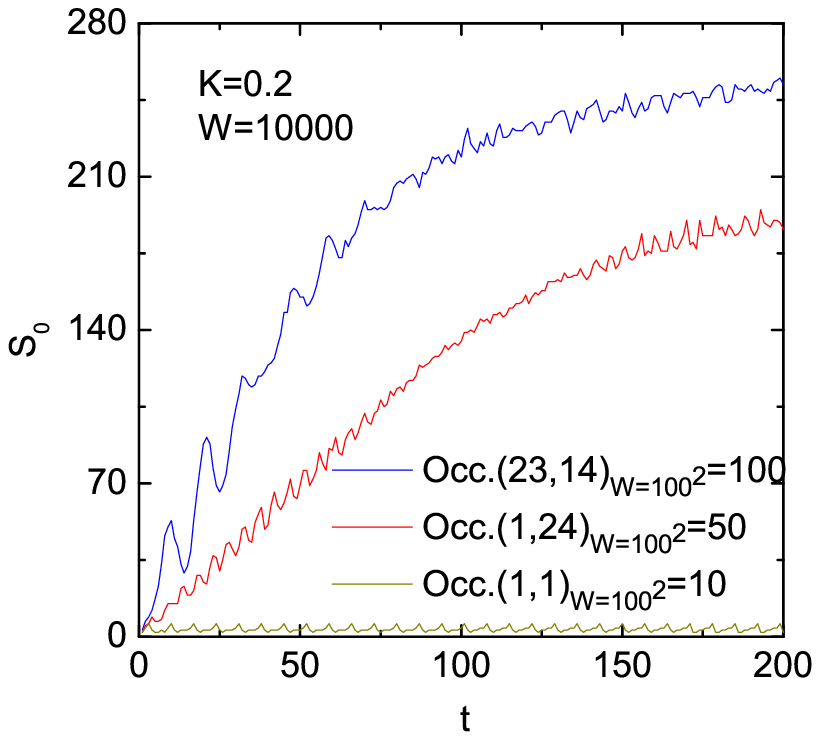}
 \caption{\label{fig:Sq0fluctuations}
(Color online)
$(q_{\text{ent}}^{\text{av}}=0)$-Entropy production of three
representative cells in a $W= 100\times 100$ equally partitioned phase space
of the standard map with $K=0.2$.
Their spreading occupation index from $t=1$ to $t=7$ are:
Occ.$(23, 14)=100$ (blue line),
Occ.$(1, 24)=50$ (red line),
Occ.$(1, 1)=10$ (brown line).
        }
\end{figure}

Entropic characterization of weak chaos regime in {\it conservative} maps
is much more subtle than in {\it dissipative} maps
(see \cite{Casati-Tsallis-Baldovin-2005, RuizTsallis}).
First, let us notice that one of the practical difficulties to identify
the weak chaos regime are the huge computer times required, because of the
slow convergence of the {\it finite time Larger Lyapunov Exponent}
to its $t \to \infty$ limit value.
Furthermore, we know that the phase space of weakly chaotic conservative maps
is dominated by stability islands and regions with a huge diversity
of spreading occupation indexes,
and this makes a hard task to reduce fluctuations by averaging the entropy.
We have already pointed out that spreading occupation index of cells,
and consequently their transient time scales, are essential
in the $q$-sensitivity characterization of the map.
Fig.\ \ref{fig:Sq0fluctuations} reveals the incidence of this phenomena on the
entropy production, which is related to the sensibility to initial conditions,
and we conclude that a smart averaging criterion must be adopted.

\begin{figure}[ht]
\centering
 \hspace{-1.5cm}
\includegraphics[width=0.39\columnwidth,keepaspectratio,clip=]{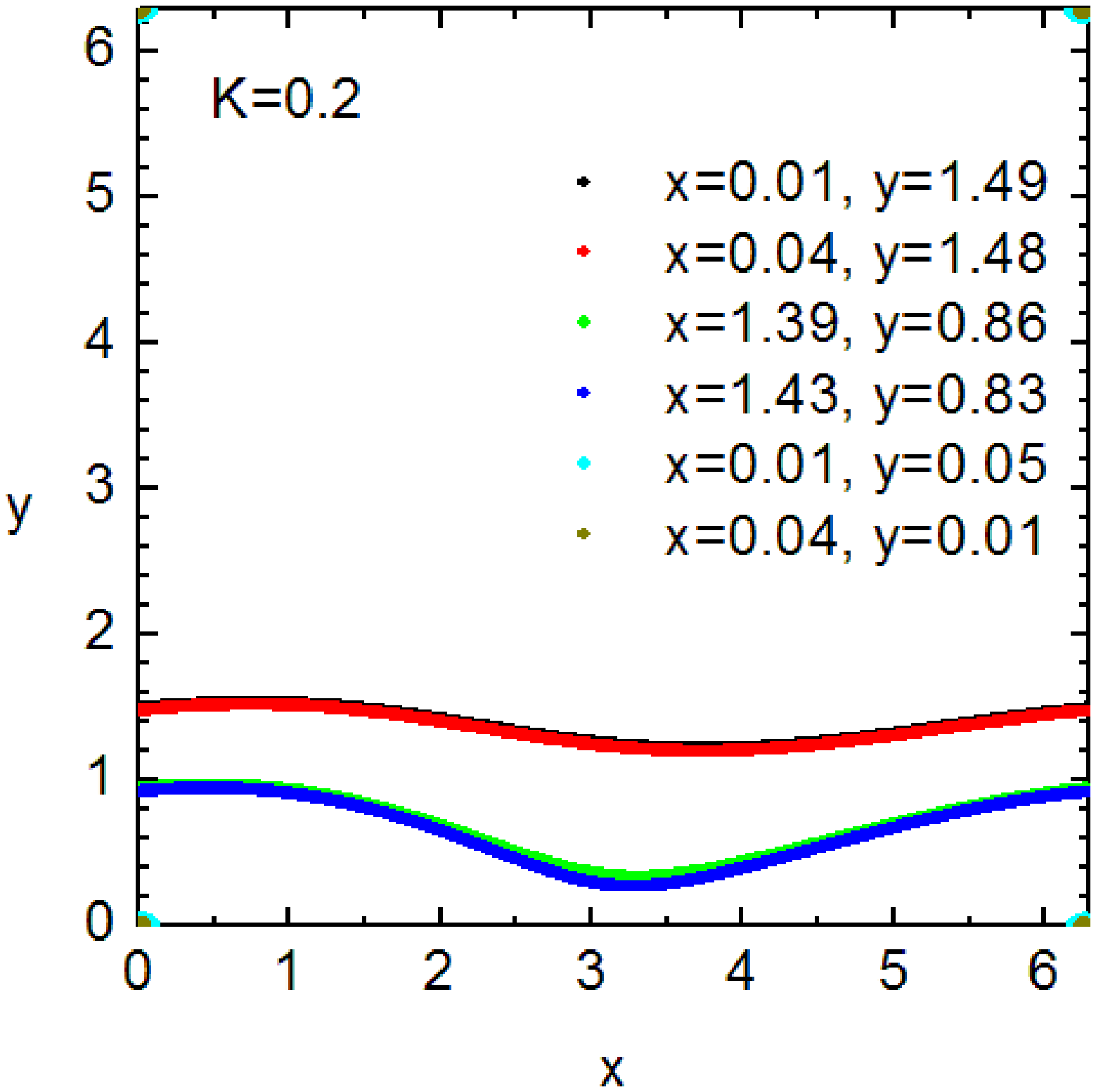}
 \hspace{-1cm}
\includegraphics[width=0.39\columnwidth,keepaspectratio,clip=]{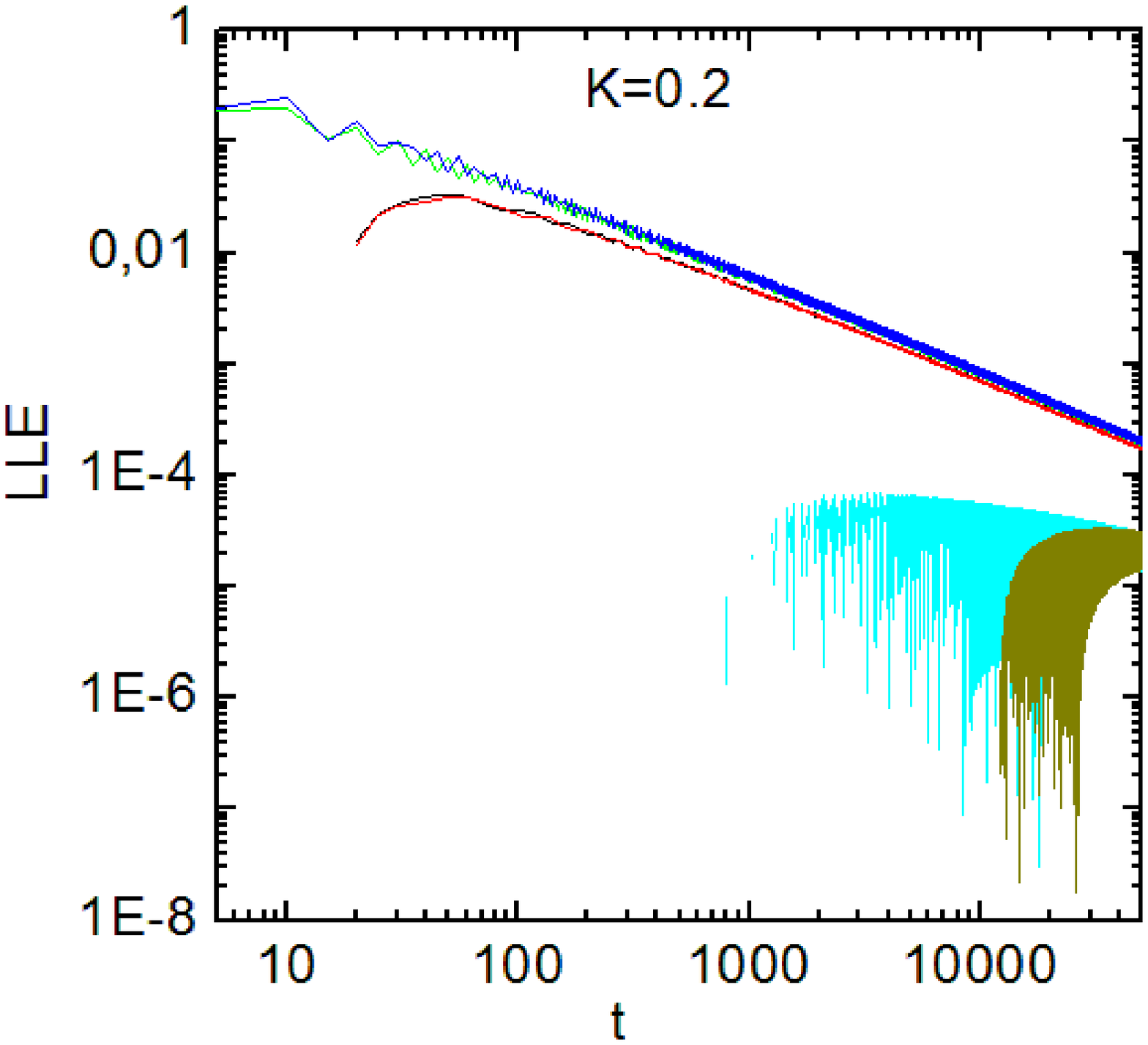}
 \hspace{-1cm}
\includegraphics[width=0.39\columnwidth,keepaspectratio,clip=]{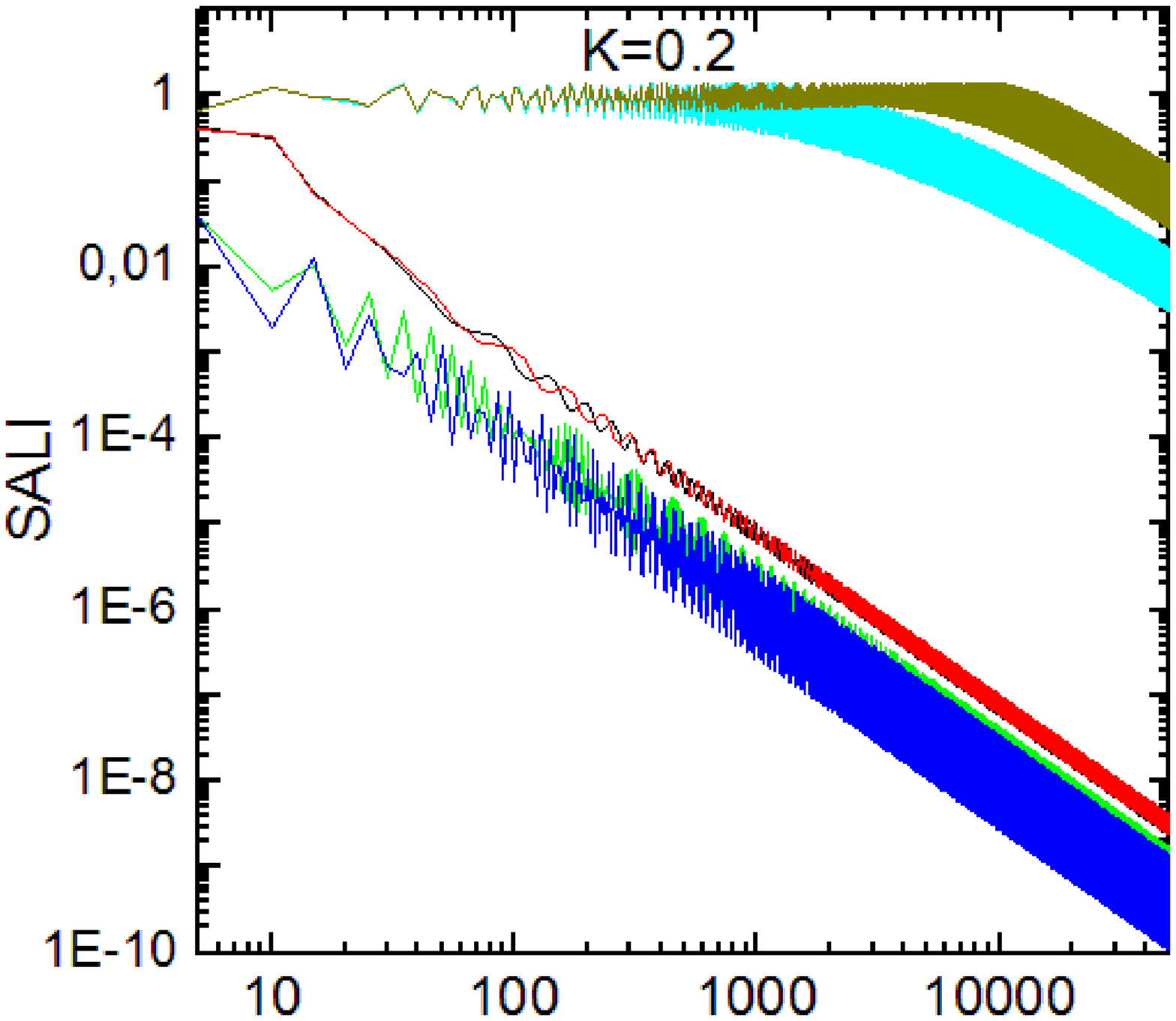}
 \hspace{-0.5cm}
 \caption{\label{fig:salvslyap}
(Color online) {\bf Left:} Paradigmatic stability island in phase space
for the standard map with $K=0.2$,
 whose respective initial conditions are specified
in the label.
{\bf Center:} Log-log representation of the respective finite time
Larger Lyapunov Exponent.
{\bf Right:} Log-log representation of the respective SALI temporal evolution.
The quasi-periodicity of trajectories are detected both by the
$t\to \infty$ limit of LLE, and by the power-law behavior
of the SALI temporal evolution.
  }
\end{figure}

\begin{figure}[ht]
\centering
 \includegraphics[width=0.5\columnwidth,keepaspectratio,clip=]{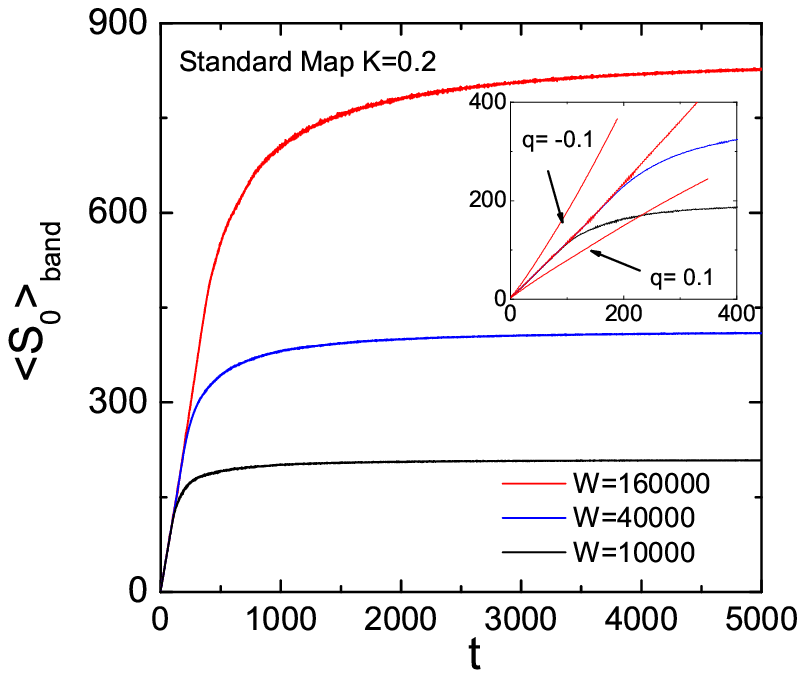}
 \caption{\label{fig:sqk02}
(Color online) Increase of $(q_{\text{ent}}^{\text{av}}=0)$-entropy
in the standard map with $K=0.2$,
for different values of $W$, all over a longitudinal band of phase space
that covers $p=2$ ($x\in [0,2\pi]$).
The inset shows that $q<q_{\text{ent}}=0$ ($q>q_{\text{ent}}=0$)
corresponds to positive (negative) concavity.
 }
\end{figure}

Initially we have tested the adequacy of the {\it Smaller Alignment Index}
(SALI) \cite{Skokos} to identify weak chaos in the $K=0.2$ standard map,
so as to reduce the  computer times required for LLE computation.
We observe that the SALI temporal evolution is a computationally cheap tool
to detect chaotic behavior of trajectories, as it tends to zero in both ordered
and chaotic orbits, but with completely different time rates:
SALI decreases abruptly in strongly chaotic orbits and reaches the limit of
accuracy of the computer after about 200 iterations and, in contrast,
it decreases with time as a power law in ordered orbits.
We have verified, for the first time to the best of our knowledge,
that the SALI decreases with time as a power law
even in the case of weakly chaotic orbits.
Fig.\ \ref{fig:salvslyap} shows, for the standard map with $K=0.2$,
that the time required to characterize weak chaos of stability island
is generically shorter if we study the SALI temporal evolution instead of
the LLE $t \to \infty$ limit computation.
We then obtain the temporal production of the averaged entropy
$\langle S_q\rangle_{N_c}$, selecting the $N_c$ cells between those
that present analogous SALI temporal evolution:
the smoother the  log-log SALI evolution is, the smother its entropy production.
This criterion
provides averaged $q$-entropies, $\langle S_q\rangle_{N_c}$,
that do not present the typical oscillations observed in conservative maps
\cite{RuizTsallis}.
Fig. \ref{fig:sqk02} exemplifies this fact.

Summarizing, for weakly chaotic dissipative systems,
particularly for the logistic map,
it has been observed that there is a specific value of $q<1$
(for the logistic map, $q_{\text{ent}}=0.2445$
\cite{latora-baranger-tsallis-rapisarda-2000})
which yields finite entropy production per unit time,
and hence $S_{q_{\text{ent}}}$ ($q_{\text{ent}} < 1$)
is the proper one to thermodynamically characterize the system.
Verification of this behavior for a conservative system (the standard map)
is presently done.
Our numerical results show in both strongly and weakly chaotic regime,
as expected, $q_{\text{ent}}^{\text{av}}=q_{\text{sen}}^{\text{av}}$,
and this points towards the conjecture of the generalized Pesin equality
\cite{tsallis-plastino-zheng-1997}.
In particular,  $q_{\text{sen}}^{\text{av}}=q_{\text{ent}}^{\text{av}}=1$
and $K_{q_{\text{ent}}^{\text{av}}}=\lambda_{q_{\text{sen}}^{\text{av}}}=1.62$
for $K=10$ (strong chaotic regime, as $\lambda>0$),
and $q_{\text{sen}}^{\text{av}}=q_{\text{ent}}^{\text{av}}=0$ for $K=0.2$
(weakly chaotic regime,
as $\lambda=0$, $\lambda_{q_{\text{sen}}^{\text{av}}} > 0$).
With respect to the weakly chaotic regime, a local characterization
of generalized Pesin-like remains to be made,
as $K_{q_{\text{ent}}^{\text{av}}}$ and $\lambda_{q_{\text{sen}}^{\text{av}}}$
have both shown to be local properties.
%

%

\section{\label{sec:q-rel} The rate of relaxation of the entropy}

The methods described in
Sections \ref{sec:q-sen} and \ref{sec:q-ent}
take an ensemble of initial conditions
inside one of $W$ cells of the phase space, and
thus $S_q(0)=0$, $\forall q$.
In Ref.\ \cite{moura-tirnakli-lyra-2000}, the authors have
considered an ensemble of initial conditions spread over the entire
phase space of a dissipative system (particularly, the logistic map),
therefore, with the maximum possible value for the entropy
$S_{q_{\text{ent}}}(0)=\ln_{q_{\text{ent}}}W$.
The time evolution brings the system to its attractor,
and the number of occupied cells falls according to an exponential
relation, in the case of a strongly chaotic system,
or according to a $q$-exponential, in the case of a weakly
chaotic system.
Later it was found a relation between  $q_{\text{rel}}$ and
$q_{\text{ent}}$ by analyzing the rate in which the entropy
goes to its final value,
beginning with $S_{q_{\text{ent}}}(0)=0$,
instead of
$S_{q_{\text{ent}}}(0)=\ln_{q_{\text{ent}}}W$
\cite{ananos-borges-tsallis-oliveira-2002}.

The method adopted in \cite{moura-tirnakli-lyra-2000}
cannot be identically applied to a conservative map,
since there is no attractor,
but that one introduced in \cite{ananos-borges-tsallis-oliveira-2002}
can be adapted to area-preserving models.
We consider the variable
 $\Delta S_q (t) = S_q(\infty) -  S_q(t)$
and evaluate its average over an ensemble of initial conditions
for $t \to \infty$.
The average must be computed with a different criterion in strong and weak chaos
regimes, according to the respective stability distributions on the phase space.

The strongly chaotic map is dominated all over the partitioned phase space
by a chaotic sea, and we can analytically calculate
$S_{q=1}(\infty) = \ln W$.
In that case, the average
$\langle \Delta S_q(t) \rangle_{N_{\text{ic}}}$ has been estimated
over the most quickly spreading cells $N_{\text{ic}}$
to minimize the time computation.
On the contrary, the stability islands dominate the phase space in the weakly
chaotic regime, and their respective value of
$S_{q_{\text{ent}}}(\infty) = \ln_{q_{\text{ent}}}W_{\text{islands}}$
is unknown, as it numerically depends on the cells where the initial conditions
are randomly chosen.
We  must find a proper criterion for the numerical estimation
of $S_{q_{\text{ent}}}(\infty)$ and
$ \langle \Delta S_q (t) \rangle_{W_{\text{islands}}}$,
and we privilege those cells that cover a region were SALI behaves
in a smoothly manner, such as to produce entropy in a smoothly manner too.

\begin{figure}[ht]
\centering
 \hspace{-1cm}
 \includegraphics[width=0.5\columnwidth,keepaspectratio,clip=]{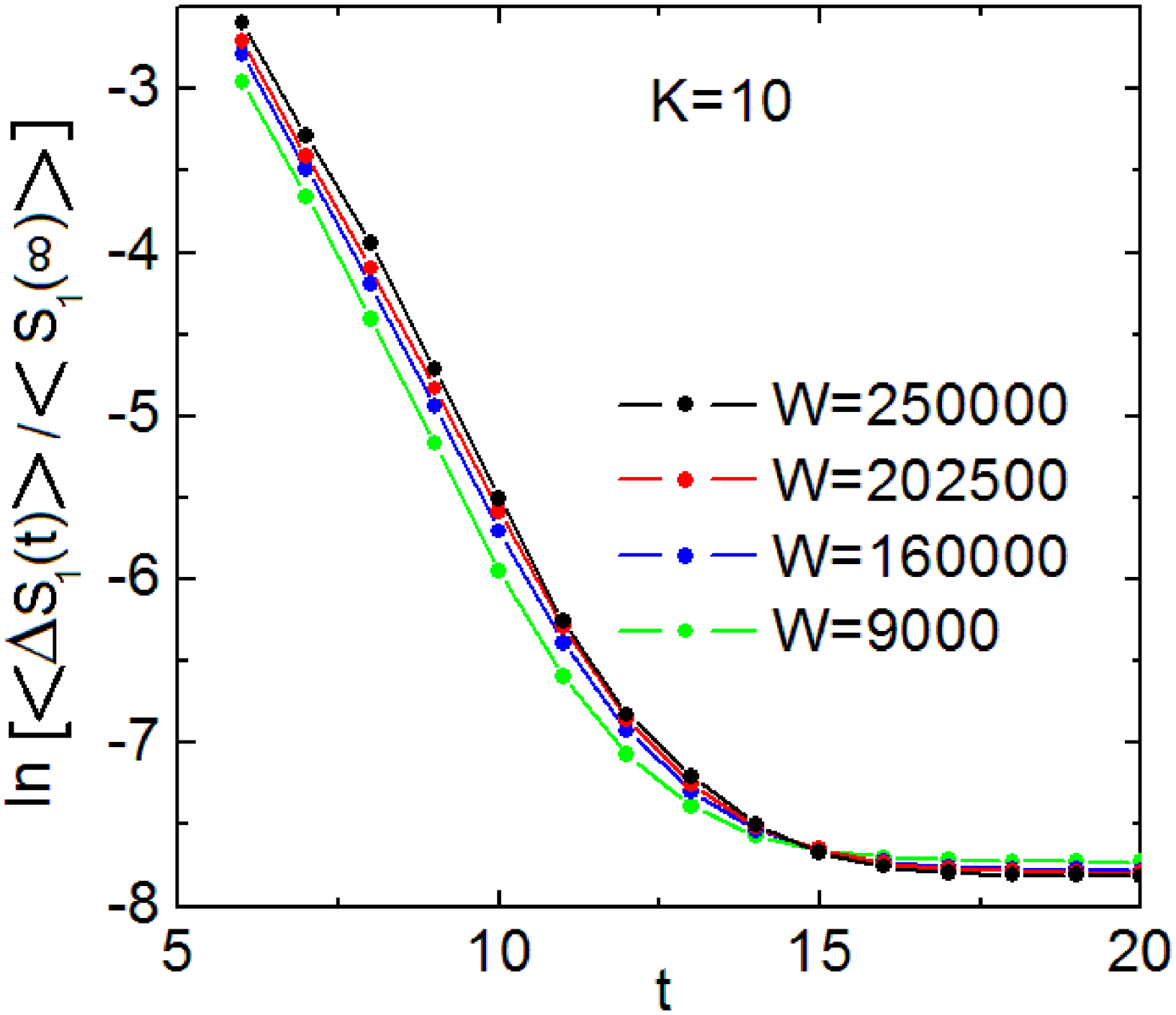}
 \hspace{-0.5cm}
 \includegraphics[width=0.55\columnwidth,keepaspectratio,clip=]{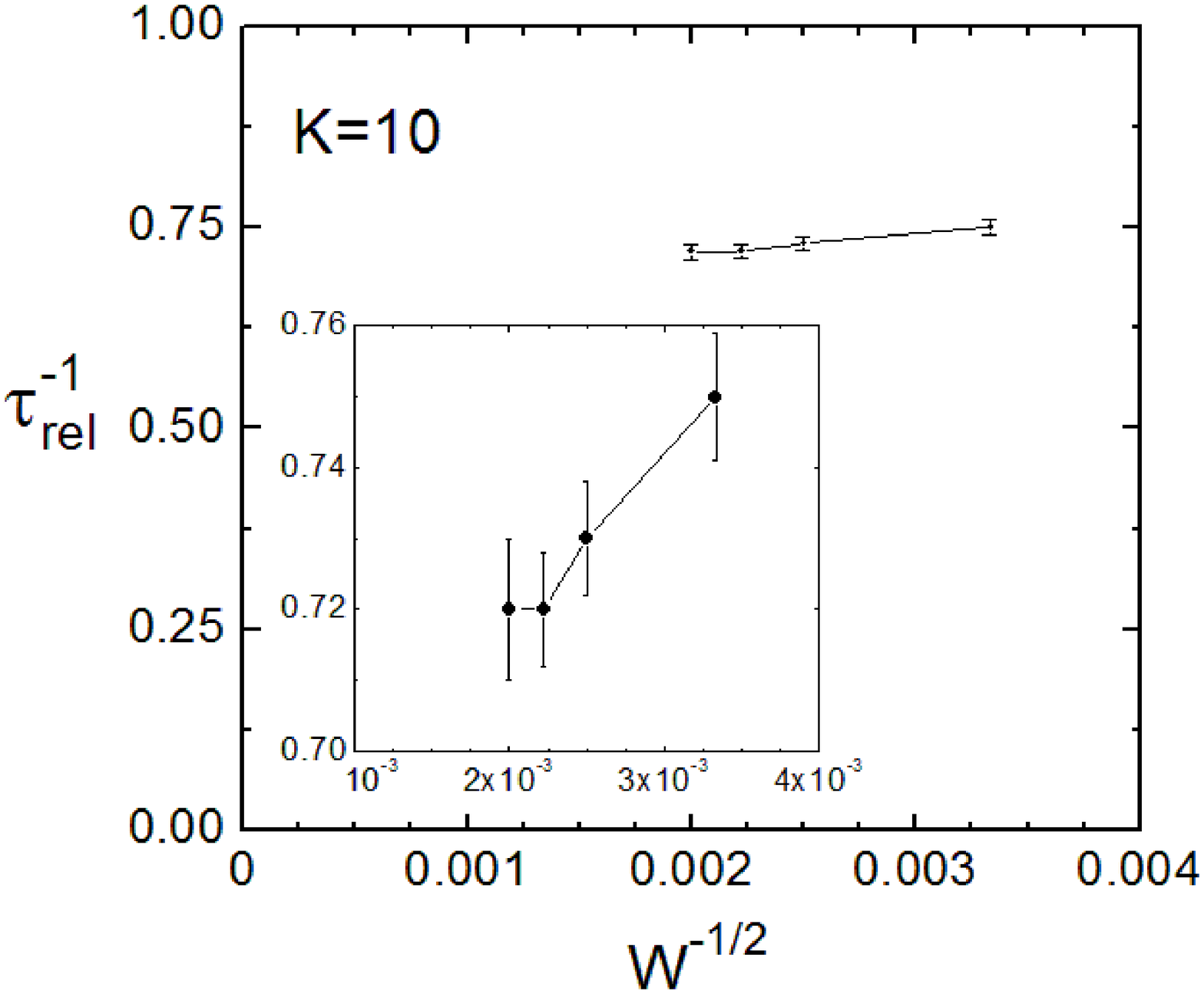}
 \caption{\label{fig:relax-k10}
(Color online) {\bf Left:} Time evolution of
$\ln_{q_{\text{rel}}}
                     \langle \Delta S_{q=1}(t) \rangle /
                     \langle S_{q=1}(\infty)   \rangle $
in the (strongly chaotic) standard map with  $K=10$.
The averages have been calculated over the more quickly spreading cells,
to optimize the computational cost.
{\bf Right:} The numerical results of the inverse of relaxation time
estimations, $\tau^{-1}_{\text{rel}} (W^{-1/2})$,
to the $S_{q_{\text{ent}}=1}(\infty)$ entropy limit, are compatible with a
finite relaxation time $\tau_{q=1}\equiv\tau_{\text{rel}}(W\to \infty)$.
}
\end{figure}

\begin{figure}[ht]
\centering
 \includegraphics[width=0.5\columnwidth,keepaspectratio,clip=]{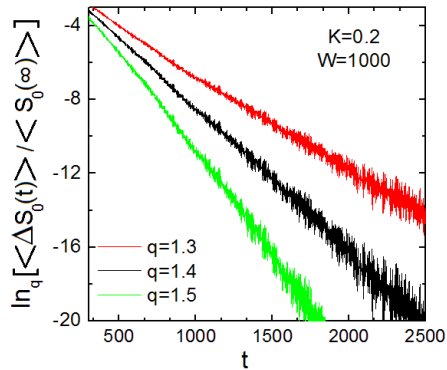}
 \caption{\label{fig:relax-k02}
(Color online)
Time evolution of
$\ln_{q_{\text{rel}}}
                     \langle \Delta S_{q=0}(t) \rangle /
                     \langle S_{q=0}(\infty)   \rangle $
in the $K=0.2$ (weakly chaotic) standard map.
The averages are calculated over the cells that belong to the same
longitudinal band in phase space analyzed in Fig. \ref{fig:sqk02}.
According to a procedure similar to what was done
in Fig.\ \protect\ref{fig:lnqzik02} and \protect\ref{fig:lnqzik02qneg},
the nonlinearity measure identifies $q_{\text{rel}}=1.4$.
}
\end{figure}

It is expected that
 $\displaystyle\lim_{W \to \infty} \Delta S_{q_{\text{ent}}^{\text{av}}}
                     \sim \exp_{q_{\text{rel}}^{\text{av}}}(-t/\tau_q)$,
where $\tau_q$ is a relaxation time.
Figures \ref{fig:relax-k10} and \ref{fig:relax-k02}
show the time dependence of
$\ln_{q_{\text{rel}}}
                    \langle \Delta S_{q_{\text{ent}}}(t) \rangle /
                    \langle S_{q_{\text{ent}}}(\infty)   \rangle $
for the strongly chaotic case ($K=10$),
and the weakly chaotic case ($K=0.2$), respectively.
The former case displays an exponential relaxation,
i.e., $q_{\text{rel}}=1$, consistently with BG framework.
The weakly chaotic case presents a $q$-exponential relaxation regime,
with $q_{\text{rel}}=1.4$ for $K=0.2$.


%
\section{\label{sec:q-stat} Stationary distributions}

Probability density distributions are among the most relevant
items regarding the statistical description of dynamical systems.
Gaussian distributions are typical for ergodic and mixing systems,
for which LLE is positive
\cite{tirnakli-beck-tsallis-2007, tirnakli-borges-2016},
and $q$-Gaussian distributions are typical for weakly chaotic systems
for which LLE$\;=0$
\cite{tirnakli-beck-tsallis-2007,tirnakli-beck-tsallis-2009, tirnakli-borges-2016}
and or chaotic layers are thin and LLE$\;\approx 0$
\cite{Ruiz-Bountis-tsallis-2012}.

We define the $q$-Gaussian distribution as
\begin{equation}
 \label{q-Gauss}
P_q(x;\mu_q,\sigma_q)=A_q\sqrt{B_q}\left[1-(1-q)B_q(x-\mu_q)^2
\right]^{\frac{1}{1-q}},
\end{equation}
where $\mu_q$ is the $q$-mean value, $\sigma_q$ is the $q$-variance,
$A_q$ is the normalization factor and $B_q$ is a parameter which characterizes
the width of the distribution
\cite{prato-tsallis-1999}:

\begin{equation}
 \label{A:q}
 A_q=\left\{\begin{array}{lc}\displaystyle\frac{\Gamma\left[\frac{5-3q}{2(1-q)}\right]}{\Gamma\left[\frac{2-q}{1-q}\right]}\sqrt{\frac{1-q}{\pi}}, &q<1\\
  \displaystyle\frac{1}{\sqrt{ \pi }},&q=1\\
\displaystyle\frac{\Gamma \left[\frac{1}{q-1}\right]}{\Gamma\left[\frac{3-q}{2(q-1)}\right]}\sqrt{\frac{q-1}{\pi}}, &1<q<3
\end{array} \right.
\end{equation}
\begin{equation}
 \label{B:q}
 B_q=[(3-q)\sigma_q^2]^{-1}
\end{equation}
and where $q\to 1$ recovers the Gaussian distribution $P_1(x; \mu_1, \sigma_1)
=\frac{1}{\sigma_1 \sqrt{2 \pi }}
e^{-\frac{1}{2}\left(\frac{x-\mu_1}{\sigma_1}\right)^2} .$

Both Gaussian and $q$-Gaussian distributions have recently been observed
for the standard map \cite{tirnakli-borges-2016}.
We have analyzed the distribution of sums of iterates of the map,
for representative $K$ values, as defined by
\begin{equation}
 \label{eq:y}
 y=\sum_{i=1}^T (x_i - \langle x \rangle ),
\end{equation}
where the average $\langle \cdots \rangle$ is both a time average
over $T$ iterations and an ensemble average over $M$ initial conditions:
\begin{equation}
\label{eq:average}
 \langle x \rangle = \frac{1}{M}\frac{1}{T}\sum_{j=1}^M \sum_{i=1}^T x_i^{(j)}.
\end{equation}
The probability distributions of $y$, namely $P(y)$,
that emerge from the standard map with weakly chaotic phase space
are $q$-Gaussians, while Gaussian distributions are associated with
the strongly chaotic regime.
The scenario for intermediate values of $K$ is also quite rich.
For these cases, the phase space displays regions with points that have
positive LLE, --- the chaotic sea ---,
and regions with $\lambda = 0$,--- the stability islands.
The probability distributions of $y$ for initial conditions
taken inside the strongly chaotic regions are Gaussians,
and those distributions for initial conditions that belong
to the weakly chaotic regions are $q$-Gaussians.
One remarkable feature is that $q_{\text{stat}}=1.935$ in all cases,
regardless the value of $K$.

If the initial conditions are spread over the entire phase space,
thus the initial conditions are from both the chaotic sea
and the stability islands,
the distribution that emerges is a linear combination of a Gaussian and
a $q$-Gaussian.
Consequently, the probability distribution of the standard map,
for any arbitrary value of $K$, can be modeled as

\begin{equation}
\label{eq:superpqGauss}
 P(y)=\alpha P_q(y;\mu_q,\sigma_q)+(1-\alpha) P_1(y;\mu_1,\sigma_1).
\end{equation}
where $P_q(y)$ and $P_1(y)$ are the probability densities of $y$
for the initial conditions taken inside the weak and strong chaos regions,
respectively.

In Ref.~\cite{tirnakli-borges-2016}
it was taken the linear combination of the probability distributions
normalized to the maximum value $P(0)$,
and the physical meaning of the interpolating parameter was not clear.
Now we have taken the linear combination of the usual probability densities in
Eq.~(\ref{eq:superpqGauss}).
We find out the ratio of the areas for the strongly and weakly chaotic regions.
In order to achieve this, we randomly generate
$N_{\text{ic}}=4\times 10^6$ initial conditions
all over the phase space  and count how many of those having non-zero Lyapunov
and those having nearly zero Lyapunov.
Dividing these numbers by the total number of initial conditions,
we mimic the ratio of the areas.
We classify an initial condition belonging to a weakly chaotic region
if its Lyapunov exponent is smaller than a given threshold
$\lambda < \lambda_- = 5\times10^{-5}$ (time steps)$^{-1}$.
Correspondingly, an initial condition is considered to belong to
a strongly chaotic region whenever
$\lambda > \lambda_+ = 10^{-2}$ (time steps)$^{-1}$.
These limiting values $\lambda_-$ and $\lambda_+$
are chosen according to what is explained in \cite{tirnakli-borges-2016}:
the phase-space ratio, i.e.,
[number of points with $\lambda < \lambda_{\text{threshold}}$] /
[number of points with $\lambda > \lambda_{\text{threshold}}$]
remains almost constant for
$\lambda_- < \lambda_{\text{threshold}} < \lambda_+$.
We call
$\alpha_1=N_{\text{ic}}(\lambda>\lambda_+)/N_{\text{ic}}$, and
$\alpha_q=N_{\text{ic}}(\lambda<\lambda_-)/N_{\text{ic}}$.
There remains a small amount of initial conditions
with intermediate Lyapunov exponent values
($\lambda_- \le \lambda \le \lambda_+$)
that are not taken into account in Eq.\ (\ref{eq:superpqGauss}),
and that are possibly negligible in the macroscopic limit.
We have thus considered the approximated form of
Eq.\ (\ref{eq:superpqGauss}) as
\begin{equation}
 \label{eq:superpqGauss-approximate}
 P(y) \approx \alpha_q P_q(y;\mu_q,\sigma_q) + \alpha_1 P_1(y;\mu_1,\sigma_1).
\end{equation}
See the results of some representative $K$ values in the Table.

\begin{table}
\begin{tabular}{|c|c|c|c|c|}
\hline
\hline
\multicolumn{5}{|c|}{$A_1=0.5642...$, $A_{q=1.935}=0.3364...$} \\
\hline
$K$  & $B_1$ & $B_q$ & $\alpha_1$ & $\alpha_q$ \\
\hline
\hline
0.2  & $-$                 & 0.015 & 0      & 1 \\
\hline
0.6  & $10.7\times10^{-7}$ & 0.015 & 0.0380 & 0.9416 \\
\hline
1    & $8.00\times10^{-7}$ & 0.015 & 0.4826 & 0.5078 \\
\hline
2    & $1.06\times10^{-7}$ & 0.010 & 0.7622 & 0.2367 \\
\hline
3    & $0.40\times10^{-7}$ & 0.010 & 0.8825 & 0.1170 \\
\hline
10   & $0.52\times10^{-7}$ & $-$   & 1      & 0 \\
\hline
\end{tabular}
\caption{\label{tab:mixed_case}
The values of the parameters for representative $K$ values.}
\end{table}

\begin{figure*}[ht]
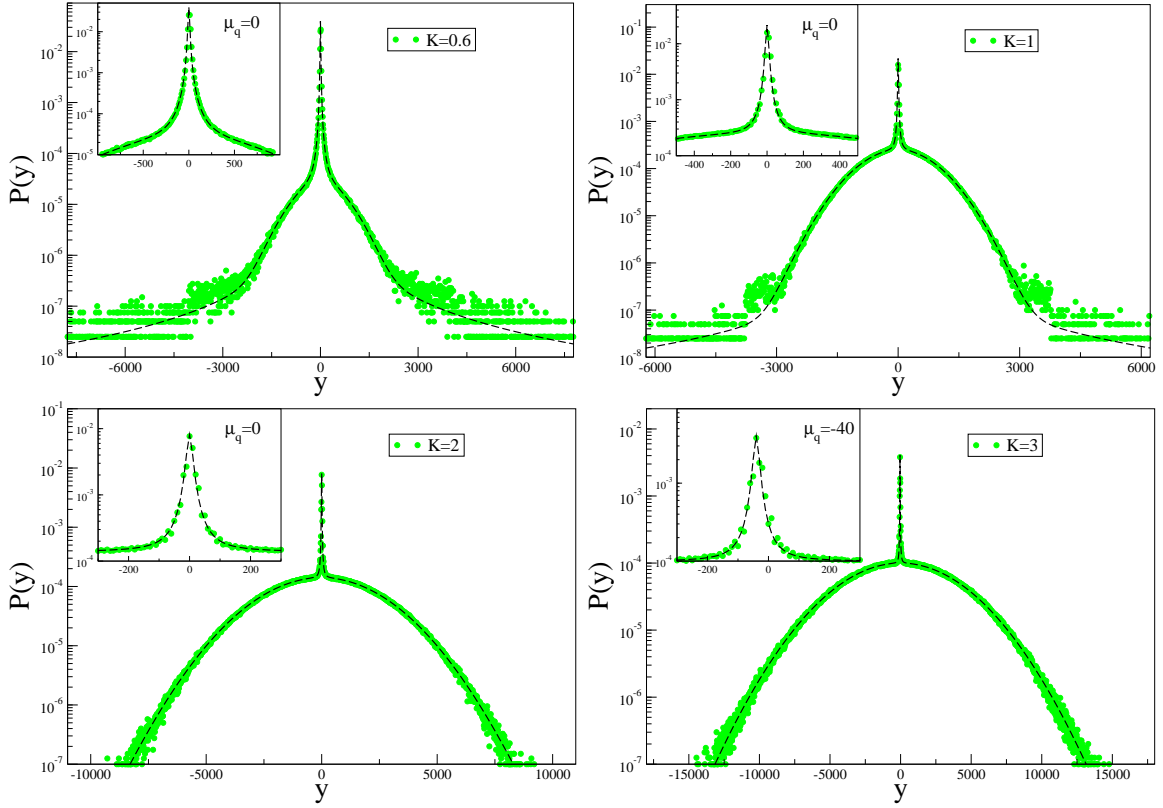

\centering
 \includegraphics[width=0.46\columnwidth,keepaspectratio,clip=]{Py-vs-y-K06.eps}
 \includegraphics[width=0.46\columnwidth,keepaspectratio,clip=]{Py-vs-y-K1.eps}
 \includegraphics[width=0.46\columnwidth,keepaspectratio,clip=]{Py-vs-y-K2.eps}
 \includegraphics[width=0.46\columnwidth,keepaspectratio,clip=]{Py-vs-y-K3.eps}
 \caption{\label{fig:alpha}
(Color online)
Probability distribution functions of the standard map
for 4 representative $K$ values.}
\end{figure*}

Then, we compute the $B_1$ and $B_q$ values using the pdf of the system
coming from the initial conditions with positive and with zero Lyapunov values,
respectively.  After determining these values now we have no fitting parameter
and $\alpha$ values are entirely determined by the intrinsic dynamics
of the map.
For some representative values of $K$, the results are given
in Fig.~\ref{fig:alpha}.

\section{Concluding Remarks}

The $q$-triplet ($q_{\text{ent}}$, $q_{\text{rel}}$, $q_{\text{stat}}$)
is an important feature in the statistical description of dynamical systems
that might be nonergodic and nonmixing, as a consequence of zero LLE.
These indices define
the rate of entropy production,
the rate of relaxation,
and the distribution of the stationary states.
Positive LLE leads to ergodic, mixing systems, and strong chaos,
and these indices collapse into $q=1$, within the BG framework.
Zero LLE may lead to breakdown of ergodicity,
nonmixing dynamics, and weak chaos.
The statistical description in many such circumstances appears to be associated with
the $q$-entropy, where the usual exponential and Gaussian functions turn into
more general forms, namely the $q$-exponential and the $q$-Gaussian ones.
This scenario has been previously observed and evaluated for dissipative
low-dimensional systems.

We have now considered a paradigmatic low-dimensional conservative system,
the standard map.
We have evaluated the $q$-triplet, and have verified that the sensitivity
to initial conditions index satisfies $q_{\text{sen}}=q_{\text{ent}}$,
in both strongly and weakly chaotic regimes.
Our results corroborate the expected situation, i.e.,
that if the phase space is completely strongly chaotic (LLE$>0$),
$q_{\text{ent}} = q_{\text{rel}} = q_{\text{stat}} = 1$ and
$K_{q_{\text{ent}}^{\text{av}}}=\lambda_{q_{\text{sen}}^{\text{av}}=1}=1.62>0$.
These  results are numerically compatible with a $q$-generalization
of a Pesin-like identity for ensemble averages,
extending those in \cite{ananos-tsallis-2004} to the case of conservative maps.
If the entire phase space is weakly chaotic regime ($\lambda \approx 0$),
which happens for low values of $K$, the standard map is characterized by
$q_{\text{ent}} = 0$,
$q_{\text{rel}} \simeq 1.4$, and
$q_{\text{stat}} \simeq 1.935$.
We have shown that  $\lambda_{q_{\text{sen}}^{\text{av}}}$
is a property that can be locally characterized by the finite temporal
evolution of LLE or, alternatively, by the SALI, which has demonstrated to be
a computationally cheaper tool to identify different chaotic regimes.
The phase space for intermediate values of $K$ features regions of
positive and regions of zero LLE.
Ensemble averages of initial conditions taken within
the strongly chaotic regions behave according to
$q_{\text{ent}} = q_{\text{rel}} = q_{\text{stat}} = 1$.
In contrast, ensemble averages of initial conditions taken within
the weakly chaotic regions behave according to the $q$-triplet
and, at least in what concerns the stationary distribution,
with the same values of $q_{\text{stat}}$.
This work extends what has been done in Ref.\ \cite{tirnakli-borges-2016}.
We have numerically found the proper superposition of distributions
for the mixed case (coexistence of strongly and weakly chaotic regimes)
without any additional fitting parameter.

The $q$-triplet structure plays a central role
on the statistical description of dynamical systems,
and additional works addressing other conservative and dissipative maps
shall bring further tests for the robustness of this framework,
and they are very welcome.

\section*{Acknowledgments}

This work has been partially supported by CNPq and Faperj
(Brazilian Agencies), and by TUBITAK (Turkish Agency)
under the Research Project number 115F492.
One of us (CT) also acknowledges partial financial support by
the John Templeton Foundation (USA).


\end{document}